\documentclass[11pt]{article}
\usepackage{moriond,epsfig}
\usepackage{pgf,pgfarrows,pgfnodes,pgfautomata,pgfheaps}
\pdfoutput=1

\def\be{\begin{equation}}
\def\ee{\end{equation}}
\def\bea{\begin{eqnarray}}
\def\eea{\end{eqnarray}}

\def\ee        {\ensuremath{e^+e^-}}
\def\pipi        {\ensuremath{\pi^+\pi^-}}
\def\piz        {\ensuremath{\pi^0}}
\def\pizpiz    {\ensuremath{\pi^0\pi^0}}
\def\fourpiz    {\ensuremath{\piz\piz\piz\piz}}

\def\bes       {\ensuremath{{\sf BESIII}}}

\def\jpsi       {\ensuremath{J\!/\psi}}
\def\psii       {\ensuremath{\psi^\prime}}

\def\chicj	{\ensuremath{\chi_{_{cJ}}}}
\def\chicz	{\ensuremath{\chi_{_{c0}}}}
\def\chicu	{\ensuremath{\chi_{_{c1}}}}
\def\chicd	{\ensuremath{\chi_{_{c2}}}}

\def\chiczd	{\ensuremath{\chi_{_{c0,2}}}}
\def\KS         {\ensuremath{K^0_{\scriptscriptstyle S}}}

\def\ff        {\ensuremath{\phi\phi}}
\def\oo        {\ensuremath{\omega\omega}}
\def\fo        {\ensuremath{\phi\omega}}
\def\etap      {\ensuremath{\eta'}}

\newcommand{\bfr}{\begin{flushright}}
\newcommand{\efr}{\end{flushright}}

\newcommand{\bfi}{\begin{figure}}
\newcommand{\efi}{\end{figure}}
\newcommand{\bc}{\begin{center}}
\newcommand{\ec}{\end{center}}
\newcommand{\bm}{\begin{minipage}}
\renewcommand{\em}{\end{minipage}}
\newcommand{\bi}{\begin{itemize}}
\newcommand{\ei}{\end{itemize}}
\newcommand{\bt}{\begin{tabular}}
\newcommand{\et}{\end{tabular}}

\begin{document}
\vspace*{4cm}
\title{Recent Results on Charmonium from BESIII}

\author{ M. MAGGIORA (on behalf of the BESIII collaboration)}

\address{Department of General Physics "A. Avogadro", University of Turin, \\Via Pietro Giuria 1, 10136 Torin, Italy}

\maketitle\abstracts{
We report the latest outcomes for the Charmonium system investigation on $226\times 10^6$ \jpsi~ and $106\times 10^6$ \psii~ events collected with the \bes~ detector at the BEPCII $e^+e^-$ collider. 
}

\section{Introduction}

Both BESIII and BEPCII represent a significant upgrade with respect to the BESII/BEPC experimental scenario. The spectrometer and the physics program, primarily aimed to investigate hadron spectroscopy and $\tau$-charm physics, are described elsewhere \cite{Asner:2008nq,Ablikim:2009vd}. The unprecedented BEPCII luminosities and the high BESIII performance allowed to collect data samples at \jpsi~ and \psii~ energies already significantly larger w.r.t those available in the literature; the analyses reported herewith have been performed on $226\times 10^6$ \jpsi~ and $106\times 10^6$ \psii~ events.

\section{\boldmath$\psii \to \pi^0 h_c$}

Clear signals have been observed 
(Fig. \ref{fig:hc_bes3})
for $\psii\to\pi^0 h_c$ with and without the subsequent radiative decay $h_c\to\gamma\eta_c$. The determination~\cite{Ablikim:2010rc} in the same experimental scenario of both $\mathcal{B}(\psii\to\pi^{0}h_c)=(8.4\pm1.3\pm1.0)\times10^{-4}$ and $\mathcal{B}(\psii\to\pi^{0}h_c)\times\mathcal{B}(h_c\to\gamma\eta_c)=(4.58\pm0.40\pm0.50)\times10^{-4}$ allows to access $\displaystyle \mathcal{B}(h_c\to\gamma\eta_c)=(54.3\pm6.7\pm5.2)\%$. $\displaystyle M(h_c)=3525.40\pm0.13\pm0.18$~MeV/$c^2$ and $\displaystyle
\Gamma(h_c)=0.73\pm0.45\pm0.28$~MeV ($<1.44$~MeV~at 90\% C.L.) have been determined as well. 

Our measurements for  $\mathcal{B}(\psii\to\pi^{0}h_c)$, $\mathcal{B}(h_c\to\gamma\eta_c)$ and $\displaystyle \Gamma(h_c)$ are the first experimental results for these quantities; the values obtained for $M(h_c)$ and  $\mathcal{B}(\psii\to\pi^{0}h_c)\times\mathcal{B}(h_c\to\gamma\eta_c)$
are consistent with previous CLEO results~\cite{Dobbs:2008ec} and of comparable precision. The measured $1P$ hyperfine mass splitting $\displaystyle \Delta~M_{hf} \equiv \langle M(1^3P) \rangle - M(1^1P_1) =-0.10 \pm 0.13 \pm0.18$~MeV/$c^2$ is consistent with no strong spin-spin interaction. For a detailed discussion of such results in the framework of the existing experimental evidences and theoretical predictions see ~\cite{Ablikim:2010rc}.

\section{\boldmath$\psii \to \gamma\chicj~;~\chiczd \to \pizpiz,\eta\eta$ ~ $(\eta,\piz \to \gamma\gamma)$}

\bfi[t]
 \bc
 \bm[t]{0.47\textwidth}
      \vspace{0pt}
      \pgfimage[width=\linewidth]{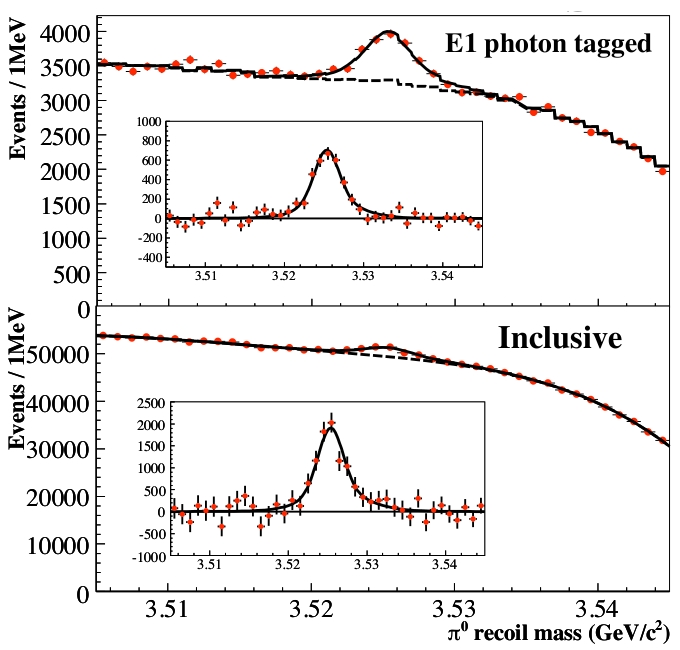}
  \em
  \ec
\caption{The $\pi^0$ recoil mass spectra and fits for: (top) the $E1$-tagged analysis: $\psii\to \pi^0h_c, h_c\to\gamma\eta_c$; (bottom) the inclusive analysis: $\psii\to\pi^{0}h_c$. Fits are shown as solid lines, background as dashed lines; insets show the background-subtracted spectra.}
\label{fig:hc_bes3}
\efi

 \bfi[!b]
   \bm[t]{0.3175\textwidth}
      \vspace{0pt}
      \pgfimage[width=\linewidth]{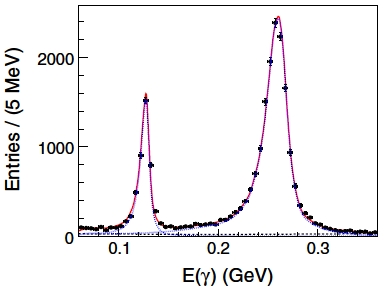}
      \setlength{\unitlength}{0.01\textwidth}
      \put(-59,48){\boldmath\color{blue}\chicd}
      \put(-25,69){\boldmath\color{blue}\chicz}
      \put(-11,67){a)}
   \end{minipage}
   \hfill
   \bm[t]{0.3175\textwidth}
      \vspace{0pt}
      \pgfimage[width=\linewidth]{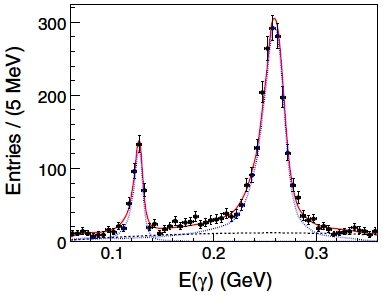}
      \setlength{\unitlength}{0.01\textwidth}
      \put(-61,40){\boldmath\color{blue}\chicd}
      \put(-25,69){\boldmath\color{blue}\chicz}
      \put(-11,68){b)}
   \em
   \hfill
   \bm[t]{0.35\textwidth}
      \vspace{-11pt}
      \pgfimage[width=\linewidth]{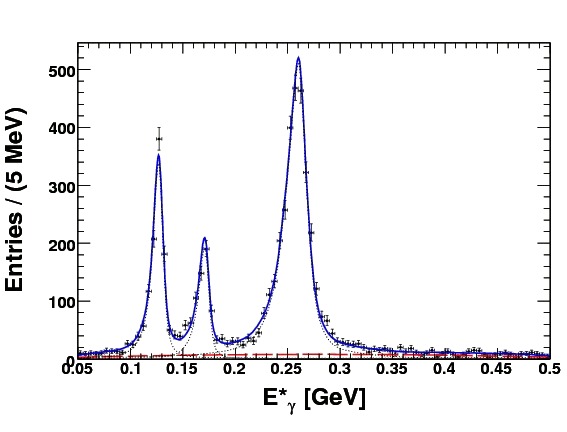}
      \setlength{\unitlength}{0.01\textwidth}
      \put(-71,54){\boldmath\color{blue}\chicd}
      \put(-63,35){\boldmath\color{blue}\chicu}
      \put(-44,63){\boldmath\color{blue}\chicz}
      \put(-11,62){c)}
   \em
\caption{The radiative photon energy spectrum of: a) selected $\chiczd\to\pi^0\pi^0$ events; b) selected $\chiczd\to\eta\eta$ events; 
c) those $\chicj\to\fourpiz$  events surviving the selection performed with the fit described in $^6$
. Fits (solid curves) include $\chi_{cJ}$ signals (dotted curves) and background polynomials (dashed curves).
}
\label{fig:chictopipi}
\efi

We have investigated the decays of the \chicz~ and \chicd~ states into the pseudo-scalar pairs $\pi^0\pi^0$ and $\eta\eta$, the corresponding \chicu~ decays being forbidden by parity conservation. Significantly clear signals (Fig. \ref{fig:chictopipi}.a-b) lead to the branching fractions: $\mathcal{B}(\chi_{c0}\to\pi^0\pi^0)=(3.23\pm 0.03\pm0.23 \pm 0.14)\times 10^{-3}$,~$\mathcal{B}(\chi_{c2}\to\pi^0\pi^0)=(8.8\pm 0.2\pm 0.6\pm0.4 )\times 10^{-4}$,~$\mathcal{B}(\chi_{c0}\to\eta\eta)=(3.44\pm 0.10\pm0.24 \pm0.2 )\times 10^{-3}$ and $\mathcal{B}(\chi_{c2}\to\eta\eta)=(6.5\pm 0.4\pm 0.5\pm 0.3)\times 10^{-4}$, where the uncertainties are statistical, systematic due to this measurement, and systematic due to the branching fractions of $\psii\to\gamma\chi_{cJ}$, respectively. For a full description of this analysis see \cite{Ablikim:2010zn}.

\section{\boldmath$\psii \to \gamma\chicj~;~\chicj \to \fourpiz$ ~ $(\piz \to \gamma\gamma)$}

The branching fractions of the $P$-wave spin-triplet Charmonium \chicj~ decays into \fourpiz~ have been determined for the first time: ${B}(\chicz\to\fourpiz)=(3.34\pm 0.06 \pm 0.44)\times10^{-3}$, ${B}(\chicu\to\fourpiz)=(0.57\pm 0.03 \pm 0.08)\times10^{-3}$ and ${B}(\chicd\to\fourpiz)=(1.21\pm 0.05 \pm 0.16)\times10^{-3}$, where the uncertainties are statistical and systematic, respectively; these fractions include decay modes with intermediate resonances except $\chicz\to\KS\KS$ and $\chicd\to\KS\KS$, which have been removed from this
measurement.  The contributions from the different states ($J=0,1,2$) are clearly visible in Fig. \ref{fig:chictopipi}.c; a complete description of the this analysis can be found in \cite{Ablikim:2010jr}.

\section{\boldmath$\chicj\to \gamma V,~V=\phi,\rho^0,\omega~;~\phi \to K^+K^-, ~\rho^0 \to \pi^+\pi^-,~\omega\to\pi^+\pi^-\pi^0$~ $(\piz \to \gamma\gamma)$}

\bfi[!t]
   \bm[t]{0.33\textwidth}
      \vspace{0pt}
      \pgfimage[width=\linewidth]{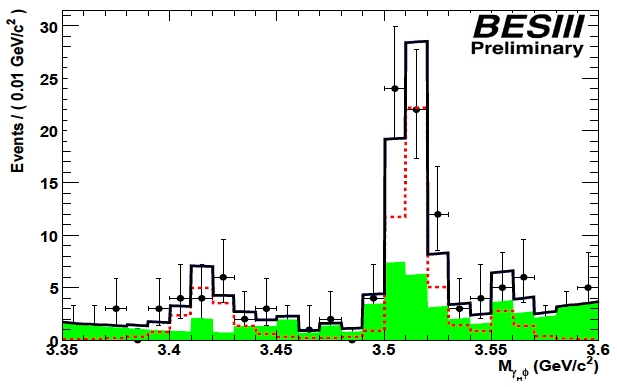}
      \setlength{\unitlength}{0.01\textwidth}
      \put(-85,54){\boldmath\color{blue}$\chicu\to\gamma\phi$}
   \end{minipage}
   \hfill
   \bm[t]{0.33\textwidth}
      \vspace{0pt}
      \pgfimage[width=\linewidth]{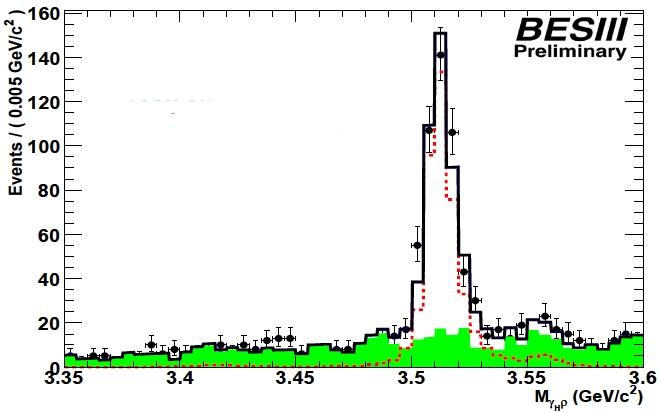}
      \setlength{\unitlength}{0.01\textwidth}
      \put(-85,54){\boldmath\color{blue}$\chicu\to\gamma\rho$}
   \em
   \hfill
   \bm[t]{0.33\textwidth}
      \vspace{0pt}
      \pgfimage[width=\linewidth]{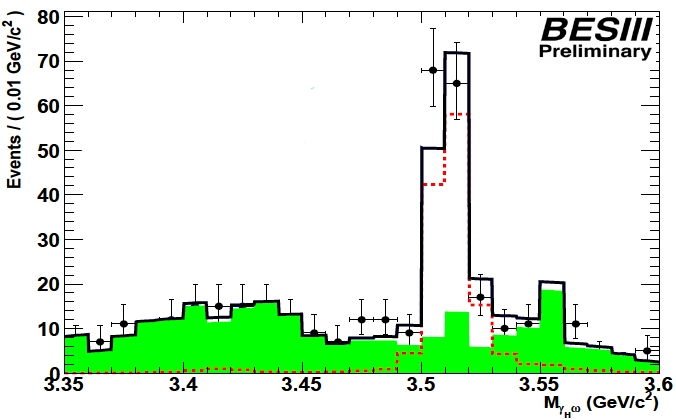}
      \setlength{\unitlength}{0.01\textwidth}
      \put(-85,54){\boldmath\color{blue}$\chicu\to\gamma\omega$}
   \em
\hfill
   \bm[t]{0.33\textwidth}
      \vspace{0pt}
      \pgfimage[width=\linewidth]{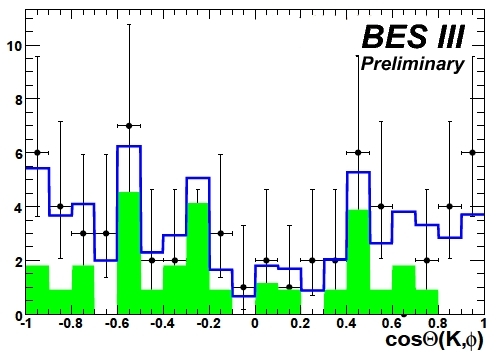}
   \end{minipage}
   \hfill
   \bm[t]{0.33\textwidth}
      \vspace{0pt}
      \pgfimage[width=\linewidth]{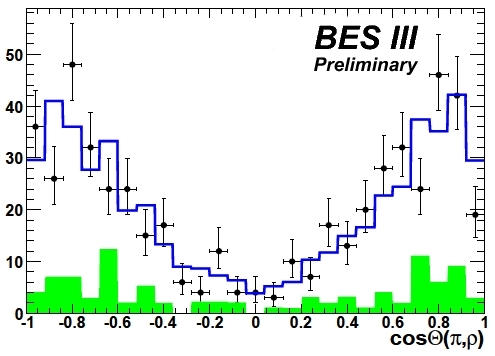}
   \em
   \hfill
   \bm[t]{0.33\textwidth}
      \vspace{0pt}
      \pgfimage[width=\linewidth]{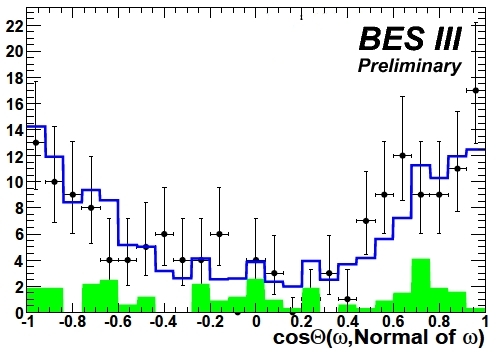}
   \em
\caption{$\chicu\to \gamma V$. Top: invariant mass distributions of (left) $K^+K^-$, (center) $\pi^+\pi^-$, and (right) $\pi^+\pi^-\pi^0$; bottom: corresponding $\cos\Theta$ distributions (see text). Histograms: best fits; dashed histograms: signal shapes; grey-shaded histograms: sum of the sideband background and the background polynomial (see $^8$).}
\label{fig:chictogammaV}
\efi

The sample of radiative $\psii\to\gamma\chicj$ events (top frames in Fig. \ref{fig:chictogammaV}) allowed to determine $\mathcal{B}(\chicu \to \gamma\rho^0)=(228\pm 13\pm 22)\times 10^{-6}$ and $\mathcal{B}(\chicu \to \gamma\omega)=(69.7\pm 7.2\pm 6.6)\times 10^{-6}$, in good agreement with earlier CLEO measurements~\cite{Bennett:2008aj}, and  
$\mathcal{B}(\chicu\to \gamma\phi)=(25.8\pm 5.2\pm 2.3)\times 10^{-6}$, observed for the first time; errors are statistical and systematic respectively. Upper limits at the 90\% confidence level on the branching fractions for $\chicz$ and $\chicd$ decays into these final states are determined as well. 
The angular dependences (bottom frames of Fig. \ref{fig:chictogammaV}) on $\cos\Theta$, $\Theta$ being the angle between the vector meson flight direction in the $\chicu$ rest frame and either the $\pi^+/K^+$ direction in the $\rho^0/\phi$ rest frame or the normal to the $\omega$ decay plane in the $\omega$ rest frame, allow to determine the fractions of the transverse polarization component of the vector meson in $\chicu\to \gamma V$ decays: $0.29_{-0.12-0.09}^{+0.13+0.10}$ for $\chi_{c1}\to \gamma\phi$, $0.158\pm 0.034^{+0.015}_{-0.014}$ for $\chi_{c1}\to \gamma\rho^0$, and $0.247_{-0.087-0.026}^{+0.090+0.044}$ for $\chi_{c1}\to \gamma\omega$. 

The present picture suggests that the longitudinal component is dominant in $\chicu\to\gamma V$ decays; for a complete description of this analysis see \cite{Ablikim:2011kv}.

\section{\boldmath$\chicj\to VV,~V=\phi,\omega~;~\phi \to K^+K^-,~\omega \to \pi^+\pi^-\pi^0$~ $(\piz \to \gamma\gamma)$}

\bfi[!b]
   \bm[t]{0.3175\textwidth}
      \vspace{0pt}
      \pgfimage[height=96pt]{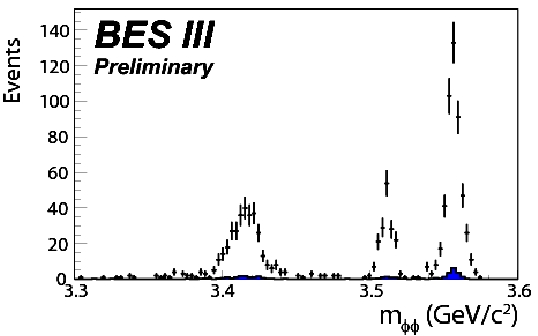}
      \setlength{\unitlength}{0.01\textwidth}
      \put(-55,25){\boldmath\color{blue}\chicz}
      \put(-30,35){\boldmath\color{blue}\chicu}
      \put(-29,57){\boldmath\color{blue}\chicd}
      \put(-11,57){a)}
   \end{minipage}
   \hfill
   \bm[t]{0.3175\textwidth}
      \vspace{0pt}
      \pgfimage[height=93pt]{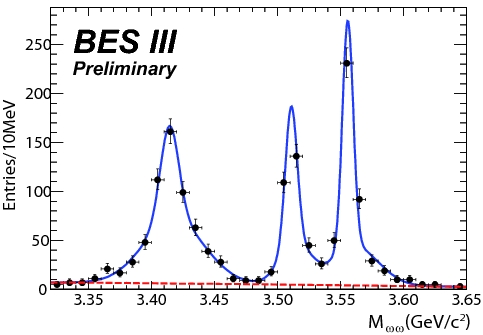}
      \setlength{\unitlength}{0.01\textwidth}
      \put(-60,44){\boldmath\color{blue}\chicz}
      \put(-39,48){\boldmath\color{blue}\chicu}
      \put(-24,57){\boldmath\color{blue}\chicd}
      \put(-11,55){b)}
   \em
   \hfill
   \bm[t]{0.35\textwidth}
      \vspace{0pt}
      \pgfimage[height=94pt]{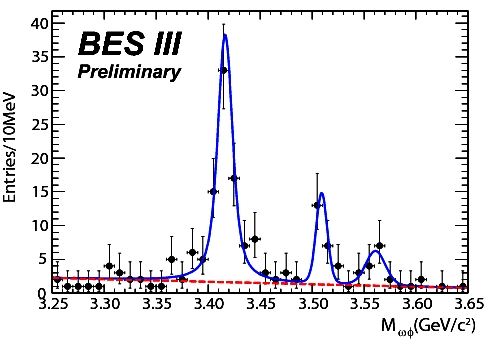}
      \setlength{\unitlength}{0.01\textwidth}
      \put(-43,50){\boldmath\color{blue}\chicz}
      \put(-34,34){\boldmath\color{blue}\chicu}
      \put(-22,25){\boldmath\color{blue}\chicd}
      \put(-11,50){c)}
   \em
\caption
{Invariant mass of a) $\ff$, b) $\oo$ and c) $\fo$. Curves: best fit results; long dash lines: fitted sideband.}
\label{fig:chictoVV}
\efi

The clear signals of Fig. \ref{fig:chictoVV} allow to investigate the $\chicj$ decays into vector meson pairs ($\ff$, $\fo$, $\oo$). 
The first observations of the $\chicu$ branching fractions $\mathcal{B}(\chicu\to\ff)=(4.4\pm0.3\pm0.5)\times10^{-4}$ and $\mathcal{B}(\chicu\to\oo)=(6.0\pm0.3\pm0.7)\times10^{-4}$ indicate that the helicity selection rule is significantly violated in these modes. 
The measured branching fractions $\mathcal{B}(\chicz\to\ff)=(8.0\pm0.3\pm0.8)\times10^{-4}$, $\mathcal{B}(\chicz\to\oo)=(9.5\pm0.3\pm1.1)\times10^{-4}$, $\mathcal{B}(\chicd\to\ff)=(10.7\pm0.3\pm1.2)\times10^{-4}$ and $\mathcal{B}(\chicd\to\oo)=(8.9\pm0.3\pm1.1)\times10^{-4}$ are consistent with and more precise than the previously published values~\cite{Nakamura:2010zzi}. 
The doubly OZI suppressed decays $\mathcal{B}(\chicz\to\fo)=(1.2\pm0.1\pm0.2)\times10^{-4}$ and $\mathcal{B}(\chicu\to\fo)=(2.2\pm0.6\pm0.2)\times10^{-5}$ are also observed for the first time. This analysis is described in details in \cite{:2011ih}.

\section{\boldmath$\psii \to \gamma P, ~ P=\piz, \eta, \eta'~;~ %\piz \to \gamma\gamma,~
  \eta \to \pi^+\pi^-\pi^0, ~ \eta \to 3\pi^0,~\eta'\to \gamma\pi^+\pi^-, ~ \eta' \to \pi^+\pi^-\eta$~ $(\eta,\piz \to \gamma\gamma)$}

\bfi[!t]
   \bm[t]{0.64\textwidth}
      \vspace{0pt}
      \pgfimage[width=\linewidth]{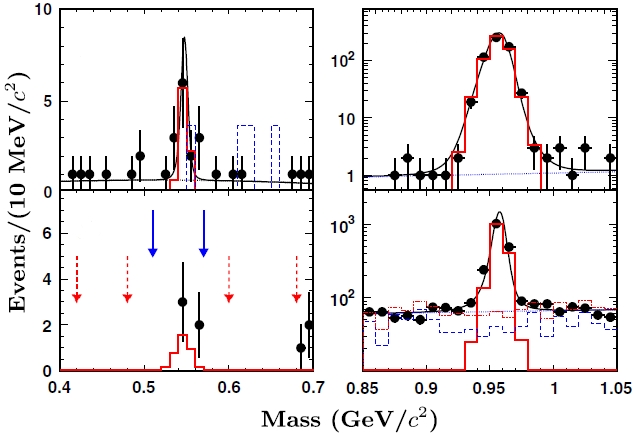}
      \setlength{\unitlength}{0.01\textwidth}
      \put(-89,63.5){\boldmath\color{blue}$\eta\to\pipi\piz$}
      \put(-41,63.5){\boldmath\color{blue}$\etap\to\gamma\pipi$}
      \put(-89,34.5){\boldmath\color{blue}$\eta\to\piz\piz\piz$}
      \put(-41,34.5){\boldmath\color{blue}$\etap\to\pipi\eta$}
      \put(-55,63.5){a)}
      \put(-55,34.5){b)}
      \put(-7,63.5){c)}
      \put(-7,34.5){d)}
   \end{minipage}
   \hfill
   \bm[t]{0.35\textwidth}
      \vspace{40pt}
      \pgfimage[width=\linewidth]{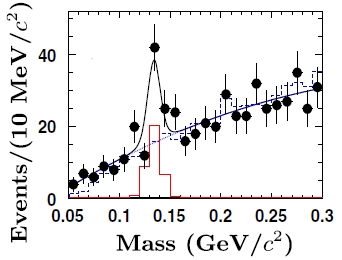}
      \setlength{\unitlength}{0.01\textwidth}
      \put(-78,66){\boldmath\color{blue}$\piz\to\gamma\gamma$}
      \put(-13,66){e)}
   \em
\caption{Mass distributions of the pseudo-scalar meson candidates for $\psii\to\gamma P$: $\gamma\eta$ [ a): $\eta\to\pipi\piz$; b): $\eta\to\piz\piz\piz$], $\gamma\etap$ [ c): $\etap\to\gamma\pipi$; d): $\etap\to\pipi\eta$, $\eta\to\gamma\gamma$], $\gamma\piz$ [ e): $\piz \to \gamma\gamma$]. 
For more details see $^{12}$.}
\label{fig:psip2gammaP}
\efi

The processes $\psii\to\gamma\pi^0$ and $\psii\to\gamma\eta$ are observed for the first time with signal significances of 4.6$\sigma$ and 4.3$\sigma$ (Fig. \ref{fig:psip2gammaP}), and branching fractions $\mathcal{B}(\psii \to\gamma\pi^0) = (1.58\pm0.40\pm0.13)\times10^{-6}$ and $\mathcal{B}(\psii\to\gamma\eta) = (1.38\pm0.48\pm0.09)\times10^{-6}$, respectively; the first errors are statistical and the second ones systematic. The branching fraction $\mathcal{B}(\psii \to\gamma\eta^\prime) = (126\pm3\pm8)\times10^{-6}$ is measured as well, leading for the first time to the determination of the ratio of the $\eta$ and $\etap$ production rates from $\psii$ decays, $R_{\psii}= \mathcal{B}(\psii\to\gamma\eta) / \mathcal{B}(\psii\to\gamma\etap)= (1.10\pm0.38\pm0.07)\%$; such ratio is below the 90\% C.L. upper bound determined by the CLEO Collaboration ~\cite{Pedlar:2009tia} and one order of magnitude smaller w.r.t the corresponding $\eta - \etap$ production ratio for the $\jpsi$ decays, $R_{\jpsi}=(21.1\pm0.9)\%$~\cite{Pedlar:2009tia}. For a detailed description of this analysis see \cite{Ablikim:2010dx}.

\end{document}